\documentclass[11pt]{article}
\usepackage{natbib,amssymb,amsmath,fullpage,hyperref}

\usepackage{Sweave}
\begin{document}

\newcommand{\m}[1]{\mathbf{#1}}
\newcommand{\bsym}[1]{\boldsymbol{#1}}
\newcommand{\etr}{{\rm etr}}

\title{Bayesian analysis of matrix data with 
              {\bf rstiefel}} 
\author{Peter Hoff \thanks{Departments of Statistics and Biostatistics, University of Washington, Seattle, WA 98195-4322. This was supported in part by
NI-CHD grant 1R01HD067509-01A1. } }
\maketitle

\begin{abstract}
We illustrate the use 
of the {\sf R}-package {\bf rstiefel}  for matrix-variate data analysis 
in the context of two examples. The first example considers 
estimation of a reduced-rank mean matrix in the presence of 
normally distributed noise. The second example considers the  modeling of a 
social network of friendships among teenagers. Bayesian estimation for 
these models 
requires the ability to simulate from the matrix-variate  
von Mises-Fisher distributions  and the matrix-variate Bingham distributions
on the Stiefel manifold. 
\end{abstract}

\section{Exponential families on the Stiefel manifold}
The set of $m\times R$ matrices $\m U$ for which $\m U^T\m U = \m I_R$ 
is called the $m\times R$ Stiefel manifold and is denoted $\mathcal V_{R,m}$. 
The densities of a 
quadratic exponential family on this manifold (with respect to the uniform measure) are given by 
\begin{equation}
p(\m U | \m A,\m B, \m C) \propto \etr ( \m C^T \m U + \m B \m U^T \m A \m U) ,
\label{eq:bmfmat}
\end{equation}
where 
$\m C \in \mathbb R^{m\times R}$, 
$\m B$ is an $R\times R$ diagonal matrix and 
$\m A$ is a symmetric matrix.  Since 
$\m U^T\m U = \m I$,  the density is unchanged under transformations of the form $\m A \rightarrow \m A + a\m I$ or $\m B \rightarrow \m B + b\m I$. 
Additionally, it is convenient to restrict the diagonal entries 
of $\m B$ to be in decreasing order. If $\m B$ is not ordered in this way, 
there exists  a reparameterization $(\m A, \tilde {\m B }, \tilde {\m C})$
giving the same distribution 
as $(\m A, \m B,\m C)$
but where $\tilde{\m B}$ has ordered diagonal
entries. 
More details on the Stiefel manifold and these distributions 
can be found in 
\citet{chikuse_2003a}, \citet{hoff_2009a}, \citet{hoff_2009b} and 
the references therein. 

Distributions of this form were originally studied 
in the case 
$R=1$, so that the manifold was just the surface of the $m$-sphere. 
In this case, $\m B$ reduces to a scalar and can be absorbed into 
the matrix $\m A$. The quadratic exponential family then has densities 
of the form 
\begin{equation}
p(\m u | \m c, \m A) \propto \exp ( \m c^T\m u + \m u^T\m A \m u). 
\label{eq:bmfvec}
\end{equation}
The case that $\m A = \m 0$ was studied by 
 von Mises, Fisher and Langevin, 
and so a distribution with density proportional to $\exp(\m c^T\m u)$ 
is often called a von Mises-Fisher or Langevin distribution 
on the sphere. 
The case that 
 $\m c= \m 0$ and $\m A \neq 0$ was studied by 
\citet{bingham_1974}, and is called the Bingham distribution. 
This distribution 
has ``antipodal symmetry'' in that $p(\m u | \m A) = p(-\m u |\m A)$, 
and so may be appropriate 
as a model for random axes, rather than random directions. 

In recognition of the work of the above mentioned authors, we
refer to distributions with densities given 
by (\ref{eq:bmfvec}) and (\ref{eq:bmfmat}) as 
vector-variate and matrix-variate 
Bingham-von Mises-Fisher distributions, respectively. 
This is a rather long name, however, so in this vignette
I will refer to them as BMF distributions.  
The case that $\m A$ (or $\m B$) is the zero matrix will 
be referred to as an MF distribution, and the case that 
$\m C$ is zero will be referred to as a Bingham distribution. 
More descriptive names might be L, Q and LQ to replace the names 
MF, Bingham, and BMF, respectively, 
the idea being that the ``L'' and ``Q'' refer to 
the presence of linear and quadratic components of the density. 

\section{Model-based SVD}
It is often useful to model an 
 $m \times n$  
rectangular matrix-variate dataset $\m Y$ 
 as being equal to 
some reduced rank matrix $\m M$ plus i.i.d.\ noise, so 
that $\m Y = \m M + \m E $, with 
the elements $\{\epsilon_{i,j}: 1\leq i\leq m, 
 1\leq  j\leq n  \}$ of $\m E$ assumed to be 
i.i.d.\ with zero mean and some unknown variance $\sigma^2$. 
The singular value decomposition states that any rank-$R$ 
matrix  $\m M$ 
can be expressed as $\m M = \m U \m D \m V^T$, where 
$\m U \in \mathcal V_{R,m}$, $\m V \in \mathcal V_{R,n}$ and 
$\m D$ is an $R\times R $  diagonal matrix.
If we are willing to assume normality of the errors, 
the model can then be written as 
\begin{eqnarray*}
\m Y &=& \m U \m D \m V^T + \m E \\
\m E& = &\{\epsilon_{i,j}: 1\leq i\leq m, 
 1\leq  j\leq n  \}  \sim \mbox{ i.i.d.\ normal}(0,\sigma^2) . 
\end{eqnarray*}
Bayesian rank selection for this model was considered in \citet{hoff_2007b}. 
In this vignette we consider estimation for a specified rank $R$, in which case 
the unknown parameters in the model are $\{ \m U , \m D, \m V , \sigma^2\}$. 
Given a suitable prior distribution over these parameters, 
Bayesian inference can proceed via construction of a Markov chain 
with stationary distribution equal to the conditional distribution of 
the parameters given $\m Y$, i.e.\ the distribution with density
$p( \m U , \m D, \m V , \sigma^2 | \m Y )$. 
In particular, conjugate prior distributions allow the 
construction of a Markov chain via the Gibbs sampler,  which  iteratively 
simulates each parameter from its full conditional distribution. 
If the prior distribution for $\m U$ is uniform on $\mathcal V_{R,m}$, 
then its full conditional density is given by 
\begin{eqnarray*}
p(\m U |  \m Y, \m D, \m V , \sigma^2 ) &\propto &  
  p(\m Y | \m U , \m D, \m V , \sigma^2  ) \\
&\propto& \etr(-[\m Y-\m U\m D\m V^T]^T[ \m Y-\m U\m D\m V^T ]/(2\sigma^2))  \\
&\propto&  \etr( [ \m Y \m V \m D/\sigma^2 ]^T \m U ) , 
\end{eqnarray*}
which is the density of an  MF$( \m Y \m V \m D/\sigma^2 )$ distribution. 
Similarly, the full conditional distribution of $\m V$  under 
a uniform prior 
is  MF$( \m Y^T \m U \m D/\sigma^2 )$. 
For this vignette, we will use the following 
prior distributions for  $\{d_1,\ldots, d_R,\sigma^2\}$:
\begin{eqnarray*}
\{ d_1,\ldots, d_R|\tau^2 \} & \sim  & \mbox{ i.i.d.\  normal}(0,\tau^2) \\
1/\tau^2 &\sim & {\rm gamma}(\eta_0/2, \eta_0\tau_0^2/2 ) \\
1/\sigma^2 &\sim &  {\rm gamma}(\nu_0/2, \nu_0\sigma_0^2/2. ) 
\end{eqnarray*}
The corresponding full conditional distributions are 
\begin{eqnarray*}
\{ d_j| \m U , \m V, \m Y , \m d_{-j}, \sigma^2 ,\tau^2\} & \sim  & \mbox{normal}(\tau^2 \m u_j^T\m Y\m v_j/[\sigma^2+\tau^2],\tau^2\sigma^2/[\tau^2+ \sigma^2] ) \\
\{1/\tau^2 |  \m U , \m D , \m V , \m Y , \sigma^2 \} &\sim& 
{\rm gamma}(  [\eta_0+R]/2 , [\eta_0\tau^2_0 + \sum d_j^2 ]/2 )
 \\
 \{1/\sigma^2 | \m U , \m D , \m V , \m Y , \tau^2 \} &\sim&  
 {\rm gamma}( [\nu_0+m n]/2 , [\nu_0\sigma^2_0 + || \m Y -\m U\m D\m V^T||^2]/2). 
\end{eqnarray*}

\subsection{Simulated data}
We now randomly generate some parameters and data according to the model 
above:
\begin{Schunk}
\begin{Sinput}
> library(rstiefel)
> set.seed(1)
> m<-60 ; n<-40 ; R0<-4
> U0<-rustiefel(m,R0)
> V0<-rustiefel(n,R0)
> D0<-diag(sort(rexp(R0),decreasing=TRUE))*sqrt(m*n)
> M0<-U0%*%D0%*%t(V0)
> Y<-M0 + matrix(rnorm(n*m),m,n)
\end{Sinput}
\end{Schunk}
The only command from the {\sf rstiefel} package used here is
{\tt rustiefel},  which generates a 
uniformly distributed random orthonormal 
matrix. Note that  {\tt rustiefel(m,R)}  gives a matrix with 
$m$ rows and $R$ columns, and so the arguments are in the 
 reverse of their order 
in the symbolic representation of the manifold $\mathcal V_{R,m}$. 
\subsection{Gibbs sampler}
Now we try to recover the true values of the parameters $\{\m U_0,\m V_0, \m D_0, \sigma^2\}$ from the observed data $\m Y$. 
Just for fun, let's estimate these parameters with a presumed rank $R>R_0$ 
that is larger than the actual rank. Equivalently, we can think of 
$\m U_0, \m V_0, \m D_0$ as having dimension $m\times R$, 
$n\times R$ and $R\times R$, but with the last $R-R_0$ diagonal 
entries of $\m D_0$ being zero. 

The prior distributions 
 for $\m U$ and $\m V$ are uniform on their respective manifolds. 
We set our hyperparameters for the other priors as follows:
\begin{Schunk}
\begin{Sinput}
> nu0<-1 ; s20<-1      #inverse-gamma prior for the error variance s2
> eta0<-1 ; t20<-1     #inverse-gamma prior for the variance t2 of the sing vals
\end{Sinput}
\end{Schunk}
Construction of a Gibbs sampler requires starting values for all 
(but one) of the unknown parameters. An natural choice is the MLE:

\begin{Schunk}
\begin{Sinput}
> R<-6
> tmp<-svd(Y) ; U<-tmp$u[,1:R] ; V<-tmp$v[,1:R] ; D<-diag(tmp$d[1:R]) 
> s2<-var(c(Y-U%*%D%*%t(V)))
> t2<-mean(diag(D^2))
\end{Sinput}
\end{Schunk}
Let's compare the MLE of $\m D$ to the true value: 

\begin{Schunk}
\begin{Sinput}
> d.mle<-diag(D) 
> d.mle
\end{Sinput}
\begin{Soutput}
[1] 40.05172 25.00226 19.70827 13.43382 13.10381 12.64942
\end{Soutput}
\begin{Sinput}
> diag(D0)
\end{Sinput}
\begin{Soutput}
[1] 38.514216 24.015791 17.352783  1.169442
\end{Soutput}
\end{Schunk}
The values of the MLE are, as expected, larger than the 
true values, especially for the smaller values of $\m D_0$. 
Now let's see if the Bayes estimate provides some shrinkage. 

\begin{Schunk}
\begin{Sinput}
> MPS<-matrix(0,m,n) ; DPS<-NULL
> for(s in 1:2500)
+ {
+   U<-rmf.matrix(Y%*%V%*%D/s2)
+   V<-rmf.matrix(t(Y)%*%U%*%D/s2)
+ 
+   vd<-1/(1/s2+1/t2)
+   ed<-vd*(diag(t(U)%*%Y%*%V)/s2 )
+   D<-diag(rnorm(R,ed,sqrt(vd)))
+ 
+   s2<-1/rgamma(1, (nu0+m*n)/2 , (nu0*s20 + sum((Y-U%*%D%*%t(V))^2))/2 )
+   t2<-1/rgamma(1, (eta0+R)/2, (eta0*t20 + sum(D^2))/2)
+ 
+   ### save output
+   if(s%%5==0)
+   {
+     DPS<-rbind(DPS,sort(diag(abs(D)),decreasing=TRUE))
+     M<-U%*%D%*%t(V)
+     MPS<-MPS+M
+   }
+ }
\end{Sinput}
\end{Schunk}
This generates a Gibbs sampler of 2500 iterations. Here, we save the 
values of $\m D$ every 5th iteration, resulting in a sample 
of $\m D$-values  of size 
500 with which to estimate $p(\m D | \m Y)$.  Additionally, 
we can obtain a posterior mean estimate of $\m M_0=\m U_0\m D_0 \m V_0^T$
via the sample average of $\m U\m D\m V^T$. Note that this estimate is 
not of rank $R$, as the set matrices of less than full rank is not convex. If we want a rank $R$ estimate, we could take the 
rank-$R$ approximation of the posterior mean. 

Let's look at the squared error for the 
MLE, the 
posterior expectation of 
$\m M_0$, and the rank-$R$ approximation to the posterior expectation:
\begin{Schunk}
\begin{Sinput}
> tmp<-svd(Y) ; M.ml<-tmp$u[,1:R]%*%diag(tmp$d[1:R])%*%t(tmp$v[,1:R])
> M.b1<-MPS/dim(DPS)[1]
> tmp<-svd(M.b1) ; M.b2<-tmp$u[,1:R]%*%diag(tmp$d[1:R])%*%t(tmp$v[,1:R]) 
> mean( (M0-M.ml)^2 )
\end{Sinput}
\begin{Soutput}
[1] 0.3563462
\end{Soutput}
\begin{Sinput}
> mean( (M0-M.b1)^2 )
\end{Sinput}
\begin{Soutput}
[1] 0.1315899
\end{Soutput}
\begin{Sinput}
> mean( (M0-M.b2)^2 )
\end{Sinput}
\begin{Soutput}
[1] 0.1311898
\end{Soutput}
\end{Schunk}
Not surprisingly, the MLE has a much larger loss than the Bayes estimates. 
The squared error for the two Bayes estimates are nearly identical. 
This is because although the posterior mean has  full rank $m \wedge n$, 
it is very close to its rank-$R$ approximation. 

\setkeys{Gin}{width=1\textwidth}

\begin{figure} 
\begin{center} 
\includegraphics{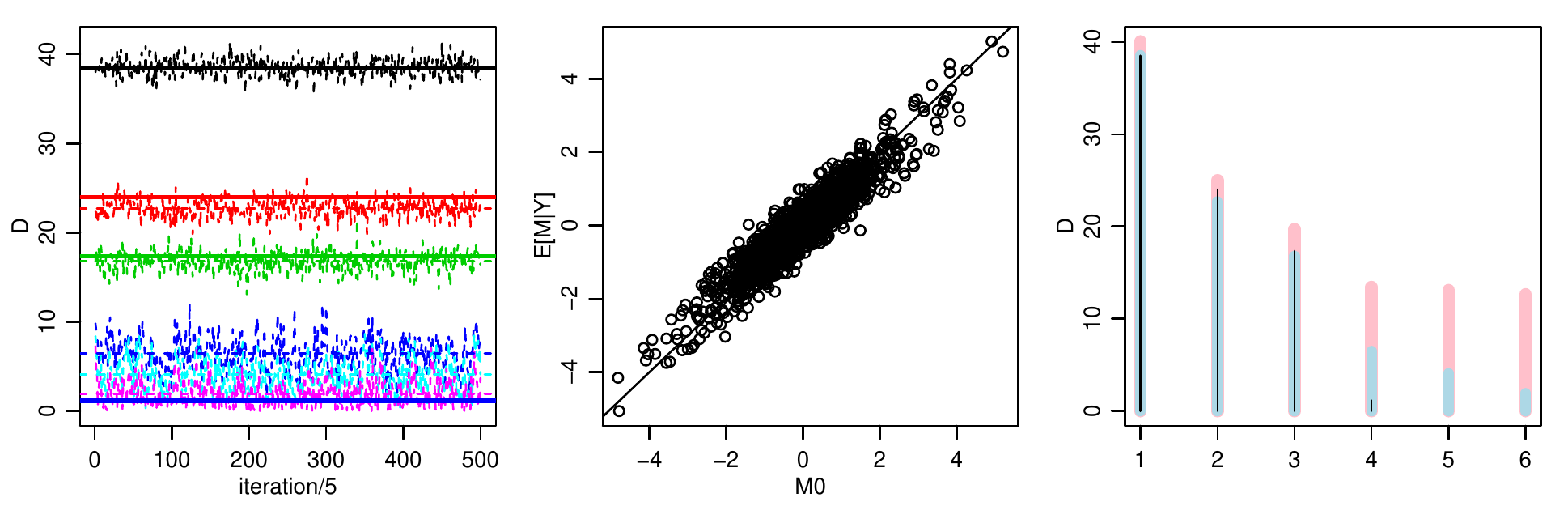}
\end{center}
\caption{Some output of the Gibbs sampler.
} 
\label{fig:svd}
\end{figure}

Finally, let's make some plots based on the output of the Gibbs sampler.
The left-most plot of Figure \ref{fig:svd}
 gives simulated values of $\m D$, with the values of $\m D_0$  
given in thick lines. 
The mixing of the Markov chain looks pretty reasonable. 
The center plot gives $\m M_0$ versus its posterior   
expectation, approximated from the MCMC sample average 
of $\m U\m D\m V^T$.  
The right plot gives the MLEs of $\m D_0$ in pink, the   
posterior expectations of $\m D_0$ in light blue, and the true values in thin black lines. The posterior estimates are very accurate for the 
large singular values of $\m D_0$, but are overestimates for the smallest 
values (the last $R-R_0$ of which are zero).  However, 
these Bayes estimates are much better than the unregularized MLEs.

\section{Network analysis}
The package {\sf rstiefel} includes a dataset on the social 
network and some health behaviors of a group of $n=50$ Scottish teenage girls. 
These data were derived from the data available at
\url{http://www.stats.ox.ac.uk/~snijders/siena/s50_data.htm}
and described in 
\citet{michell_amos_1997}.

\subsection{An eigenmodel for symmetric networks}
Let $\m Y$ be the $n\times n$ symmetric adjacency matrix 
corresponding to this network, 
with off-diagonal
 entry $y_{i,j}$ equal to the binary indicator of a friendship between 
actors $i$ and $j$, as reported by one or both actors. 
In this vignette we will
derive a model-based representation of these data 
using the following 
reduced-rank probit model:
\begin{eqnarray}
z_{i,j} &=& \theta + \m u_i^T \Lambda \m u_j + \epsilon_{i,j}  
\label{eq:eigenmodel}
\\
y_{i,j} &=& 1_{(0,\infty)}(z_{i,j}),  \nonumber
\end{eqnarray}
where 
$\{\epsilon_{i,j}=\epsilon_{j,i}\} \sim $ i.i.d.\ normal$(0,1)$, $\Lambda={\rm diag}(\lambda_1,\lambda_2)$ and the matrix $\m U$ with row vectors $\m u_1,\ldots, \m u_n$ 
lies in the Stiefel manifold $\mathcal V_{R,n}$. 
This model is a type of two-way latent factor model
in which the relationship between actors $i$ and $j$ is 
modeled in terms of their unobserved latent factors 
$\m u_i$ and $\m u_j$. This model and its relationship to other 
latent variable network models are described more fully in 
  \citet{hoff_2008}.

Convenient prior distributions for $\{ \m U , \Lambda , \theta\}$ 
are as follows:
\begin{eqnarray*}
\theta & \sim& {\rm normal}(0 ,\tau^2_\theta)  \\
(\lambda_1,\lambda_2) &\sim&  \mbox{ i.i.d.\ normal}(0,\tau^2_\lambda) \\
\m U &\sim& {\rm uniform}(\mathcal V_{R,n}) 
\end{eqnarray*}
Conditional on the observed network $\m Y$, 
posterior inference 
can proceed 
via a Gibbs sampling scheme for the unknown quantities  
$\{ \m Z, \m U , \Lambda, \theta\}$. 
Under model (\ref{eq:eigenmodel}), 
observing $y_{i,j}=0 $ or $1$ implies that $z_{i,j}$ is 
less than or greater than zero, respectively. Thus 
conditional on $\{ \m Y, \m U, \Lambda, \theta\}$, 
the distribution of $\m Z$ is that of  
a random symmetric normal matrix with mean $\theta+\m U \Lambda \m U^T$
and  independent entries that are constrained to be positive 
or negative depending on the entries of $\m Y$. 
Given $\m Z$, the full conditional distributions of 
 $\{\m U , \Lambda, \theta\}$ do not depend on $\m Y$, and 
can be obtained from the corresponding prior distributions and 
the density 
 for the matrix $\m Z$, given by
\begin{eqnarray} 
 p(\m Z | \m U,\Lambda) & \propto  &
{\rm etr}( -[\m Z - \theta \m 1 \m 1^T- \m U \Lambda \m U^T ]^T [\m Z - \theta \m 1 \m 1^T - \m U \Lambda \m U^T ]/4) \nonumber  \\
&=& {\rm \etr}( -\m E^T\m E/4 )  \times 
   {\rm \etr}( \Lambda\m U^T \m E \m U /2  )  \times 
 {\rm \etr}( -\Lambda^2/4),
\label{eq:llik}
\end{eqnarray}
where $\m E = \m Z-\theta\m 1\m 1^T$  has mean $\m U\Lambda \m U^T$ 
and off-diagonal variances of 1.
The diagonal elements of $\m E$ (and $\m Z$) have variance 2, but do not correspond
to any observed data as the diagonal of $\m Y$ is undefined. These diagonal elements are integrated over in the Markov chain Monte Carlo estimation scheme
described below.
From (\ref{eq:llik}), the full conditional distribution  of $\m U$ is 
easily seen to be a Bingham$(\m E/2,\Lambda)$  distribution. 
Full conditional distributions for the other quantities are 
available via standard calculations, and are given 
in \citet{hoff_2009a} and in the code below.

\subsection{Gibbs sampler}

The data for this example are stored as a list:
\begin{Schunk}
\begin{Sinput}
> data(YX_scots) ; Y<-YX_scots$Y ; X<-YX_scots$X
\end{Sinput}
\end{Schunk}
The $n\times 2$ matrix $\m X$ provides a binary indicator 
of drug use and smoking behavior for each actor 
during the 
period of the study. 
Understanding the relationship between these health behaviors 
and the social network can be facilitated by examining the relationship
between $\m X$ and the latent factors $\m U$ that represent the network 
via the  model given in  (\ref{eq:eigenmodel}). 

We specify the dimension of the latent factors and the values of the 
hyperparameters as follows:
\begin{Schunk}
\begin{Sinput}
> ## priors 
> R<-2 ; t2.lambda<-dim(Y)[1] ; t2.theta<-100
\end{Sinput}
\end{Schunk}
A value of $\tau^2_\lambda = n$ 
allows the 
prior magnitude of the latent factor effects 
to increase with $n$, but not as fast as the residual variance:
Letting $\m U_1$ be the first column of $\m U$, 
we have ${\rm E}[ || \lambda_1 \m U_1 \m U_1^T  ||^2 ]  = 
 {\rm  E}[ \lambda_1^2 ]  = n$. On the other hand, letting 
$\mathcal E $ be the matrix of residuals $ \{ \epsilon_{i,j} \}$ , we have 
${\rm E }[ ||\mathcal E||^2 ] = (n+1)n $.

For brevity, we consider simple, naive starting values for the 
unknown parameters:
\begin{Schunk}
\begin{Sinput}
> ## starting values
> theta<-qnorm(mean(c(Y),na.rm=TRUE))
> L<-diag(0,R)
> set.seed(1)
> U<-rustiefel(dim(Y)[1],R)
\end{Sinput}
\end{Schunk}
Better starting values could be obtained from a few iterations 
of an   EM or block coordinate descent algorithm, although these naive 
starting values are adequate for this example.

We are now ready to run the Gibbs sampler. We will store 
simulated values of $\Lambda$ and  $\theta$  in the objects 
\verb@LPS@ and \verb@TPS@, respectively. Instead of saving 
values of $\m U$, we will just compute the sum of 
$\m U \Lambda  \m U^T$ across iterations of the Markov 
chain. Dividing by the number of iterations, this sum provides 
an approximation to the posterior mean of $\m U \Lambda \m U^T$. 
A rank-$R$ eigendecomposition of the posterior mean 
can be used to provide an estimate of $\m U$.

\begin{Schunk}
\begin{Sinput}
> ## MCMC
> LPS<-TPS<-NULL ; MPS<-matrix(0,dim(Y),dim(Y))
> for(s in 1:10000)
+ {
+ 
+   Z<-rZ_fc(Y,theta+U%*%L%*%t(U))
+ 
+   E<-Z-U%*%L%*%t(U)
+   v.theta<-1/(1/t2.theta + choose(dim(Y)[1],2))
+   e.theta<-v.theta*sum(E[upper.tri(E)])
+   theta<-rnorm(1,e.theta,sqrt(v.theta))
+ 
+   E<-Z-theta
+   v.lambda<-2*t2.lambda/(2+t2.lambda)
+   e.lambda<-v.lambda*diag(t(U)%*%E%*%U/2)
+   L<-diag(rnorm(R,e.lambda,sqrt(v.lambda)))
+ 
+   U<-rbing.matrix.gibbs(E/2,L,U)
+ 
+   ## output
+   if(s>100 & s%%10==0)
+   {
+     LPS<-rbind(LPS,sort(diag(L))) ; TPS<-c(TPS,theta) ; MPS<-MPS+U%*%L%*%t(U)
+   }
+ }
\end{Sinput}
\end{Schunk}
Note that this code uses a function \verb@rZ_fc@, which simulates 
from the full conditional distribution of  $\m Z$ given $\{ \m Y, \m U, \Lambda , \theta\}$, which is that of independent 
constrained normal random variables. 
The code for this function can be obtained from the \LaTeX\  source
file for this document. 

\begin{figure}
\begin{center}
\includegraphics{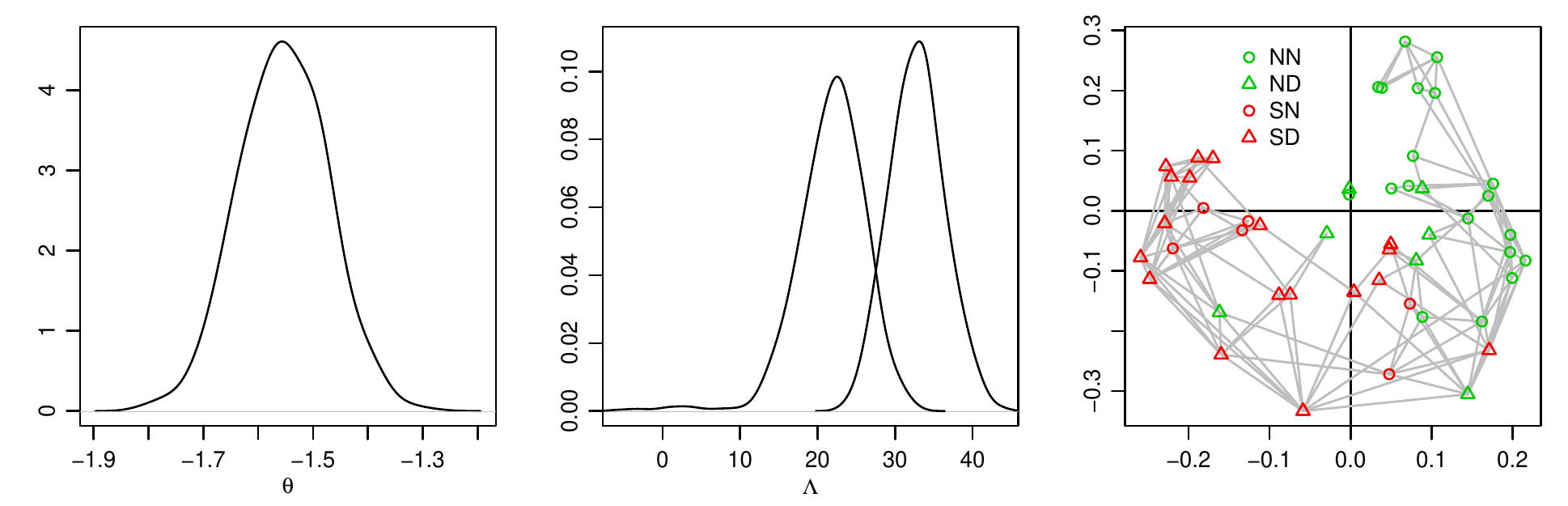}
\end{center}
\caption{Some output of the Gibbs sampler.}
\label{fig:sna}
\end{figure}

A summary of the posterior distribution is provided in 
Figure \ref{fig:sna}.  The first panel plots the posterior 
density of $\theta$, and the second plots the (marginal) posterior 
densities of the ordered values of $(\lambda_1,\lambda_2)$.  
This plot strongly suggests that the values of $\lambda_1$ and $\lambda_2$ 
 are both positive.
Since the probability of a friendship between $i$ and $j$ is increasing in $\m u_i^T \Lambda \m u_j$, 
the results posit that friendships are more likely between individuals 
with similar values for their latent factors (this effect 
is sometimes referred to as homophily). 
The third panel plots  the observed network with the node positions 
obtained from 
the estimates of 
$\m u_1,\ldots, \m u_n$ based on  the
rank-2 approximation  of the posterior mean of 
 $\m U \Lambda \m U^T$. 
The  plotting colors and characters for the nodes are determined by the 
drug and smoking behaviors: Non-smokers are plotted in green and smokers 
in red, non-drug users are plotted as circles and drug users as triangles. 
The plot indicates a separation between students with no drug or 
tobacco use (green circles) from the other students in terms of their latent factors, suggesting a relationship
between these health behaviors and the social network. 

\setkeys{Gin}{width=1\textwidth}


\end{document}